\documentclass{article}

     \usepackage[preprint,nonatbib]{neurips_2019}

\usepackage[utf8]{inputenc} 
\usepackage[T1]{fontenc}    
\usepackage{hyperref}       
\usepackage{url}            
\usepackage{amsfonts}       

\usepackage{nicefrac}       
\usepackage{microtype}      

\usepackage{booktabs}   
\usepackage{subcaption} 

\usepackage{amsmath}
\usepackage{xspace}
\usepackage{hyperref}
\usepackage{color}
\usepackage{graphics}
\usepackage{paralist}
\usepackage{listings}
\usepackage{array}
\usepackage{flushend}
\usepackage{todonotes}
\usepackage{xargs}

\newcommand{\secref}[1]{Section~\ref{sec:#1}}

\newcommand{\figref}[1]{Figure~\ref{fig:#1}}

\newcommand{\maybebreak}{}

\newcommand{\MLIR}{MLIR\xspace}

\newcommand{\Cpp}{C\texttt{++}\xspace}

\definecolor{red-dark}{rgb}{0.8,0,0}
\definecolor{green-light}{rgb}{0.2,0.5,0.2}
\definecolor{green-dark}{rgb}{0,0.3,0}
\definecolor{blue-dark}{rgb}{0,0,.8}

\def\paperversiondraft{draft}

\ifx\paperversion\paperversiondraft
\else
  \ifx\paperversion\paperversionfinal
  \else
    \errmessage{Configuration error: paperversion must be set to 'draft' or 'final'}
  \fi
\fi

\ifx\paperversion\paperversiondraft
  \paperwidth=\dimexpr \paperwidth + 6cm\relax
  \oddsidemargin=\dimexpr\oddsidemargin + 3cm\relax
  \evensidemargin=\dimexpr\evensidemargin + 3cm\relax
  \marginparwidth=\dimexpr \marginparwidth + 3cm\relax
  \newcommandx{\ac}[2][1=]{#1\todo[size=\scriptsize,linecolor=red,backgroundcolor=red!25,bordercolor=red,#1]{\textbf{Albert:} #2}}
  \newcommandx{\az}[2][1=]{#1\todo[size=\scriptsize,linecolor=red,backgroundcolor=red!25,bordercolor=red,#1]{\textbf{Alex:} #2}}
  \newcommandx{\am}[2][1=]{#1\todo[size=\scriptsize,linecolor=cyan,backgroundcolor=cyan!25,bordercolor=cyan,#1]{\textbf{Mehdi:} #2}}
  \newcommandx{\jp}[2][1=]{#1\todo[size=\scriptsize,linecolor=blue,backgroundcolor=blue!25,bordercolor=blue,#1]{\textbf{Jacques:} #2}}
  \newcommandx{\cl}[2][1=]{#1\todo[size=\scriptsize,linecolor=green,backgroundcolor=green!25,bordercolor=green,#1]{\textbf{Chris:} #2}}
  \newcommandx{\ts}[2][1=]{#1\todo[size=\scriptsize,linecolor=yellow!25!black,backgroundcolor=yellow!25,bordercolor=yellow!25!black,#1]{\textbf{Tatiana:} #2}}
  \newcommandx{\nv}[2][1=]{#1\todo[size=\scriptsize,linecolor=white,backgroundcolor=gray!25,bordercolor=blue,#1]{\textbf{Nicolas:} #2}}

  \newcommand{\TODO}[1]{{\color{red}\par\noindent\textbf{TODO:} #1\par}}
\fi

\ifx\paperversion\paperversionfinal
  \newcommand{\ac}[2][1=]{}
  \newcommand{\az}[2][1=]{}
  \newcommand{\am}[2][1=]{}
  \newcommand{\jp}[2][1=]{}
  \newcommand{\cl}[2][1=]{}
  \newcommand{\ts}[2][1=]{}
  \newcommand{\nv}[2][1=]{}

  \newcommand{\TODO}[1]{}
\fi

\lstdefinelanguage{mlir}{ 
  morekeywords={},
  sensitive=true,
  morecomment=[l]{//},
  morestring=[b]{"}
}
\lstdefinelanguage{tablegen}{ 
  morekeywords={def,class,let},
  morekeywords=[2]{Op,Pattern},
  sensitive=true,
  morecomment=[l]{//},
  morestring=[b]{"},
}
\lstdefinelanguage{ebnf}{
  alsoletter={:=|},
  morekeywords={::=,|},
  morecomment=[l]{//},
  morestring=[b]{`}
}

\title{\MLIR: A Compiler Infrastructure for the End of Moore's Law}

\author{%
  Chris Lattner
  \thanks{With SiFive at the time of publication.} \\
  Google \\
  \And
  Mehdi Amini \\
  Google \\
  \And
  Uday Bondhugula \\
  IISc \\
  \And
  Albert Cohen \\
  Google \\
  \And
  Andy Davis \\
  Google \\
  \And
  Jacques Pienaar \\
  Google \\
  \And
  River Riddle \\
  Google \\
  \And
  Tatiana Shpeisman \\
  Google \\
  \And
  Nicolas Vasilache \\
  Google \\
  \And
  Oleksandr Zinenko \\
  Google \\
}

\begin{document}

\maketitle

\begin{abstract}
  This work presents \MLIR, a novel approach to building reusable and
  extensible compiler infrastructure. \MLIR aims to address software
  fragmentation, improve compilation for heterogeneous hardware,
  significantly reduce the cost of building domain specific compilers,
  and aid in connecting existing compilers together.

  \MLIR facilitates the design and implementation of code generators,
  translators and optimizers at different levels of abstraction and
  also across application domains, hardware targets and execution
  environments. The contribution of this work includes (1) discussion
  of \MLIR as a research artifact, built for extension and evolution,
  and identifying the challenges and opportunities posed by this novel
  design point in design, semantics, optimization specification,
  system, and engineering. (2) evaluation of \MLIR as a generalized
  infrastructure that reduces the cost of building
  compilers---describing diverse use-cases to show research and
  educational opportunities for future programming languages,
  compilers, execution environments, and computer architecture. The
  paper also presents the rationale for \MLIR, its original design
  principles, structures and semantics.
\end{abstract}

\section{Introduction}\label{sec:intro}

Compiler design is a mature field with a wide range of well-known algorithms, with applications to code generation, static analysis, program transformation,
and more. The field also has seen the development of a number of mature
technology platforms which have enabled massive reuse across the compiler
community, including systems like the LLVM compiler
infrastructure~\cite{LLVM:CGO04}, the Java Virtual Machine (JVM)~\cite{Lindholm:1999:JVM:553607}, and many others.
A common characteristic of these popular systems is their ``one size fits all'' approach---a
\emph{single abstraction level} to interface with the system: the LLVM Intermediate Representation (IR) is
roughly ``C with vectors'', and JVM provides an ``object-oriented type
system with a garbage collector'' abstraction. This ``one size fits all''
approach is incredibly valuable---and in practice, the mapping to these
domains from ubiquitous source languages (C/\Cpp and Java respectively) is
straightforward.

At the same time, many problems are better modeled at a higher- or lower-level
abstraction, e.g.\ source-level analysis of \Cpp code is very difficult on LLVM
IR. We observe that many languages (including e.g.\ Swift, Rust, Julia, Fortran)
develop their own IR in order to solve domain-specific problems, like
language/library-specific optimizations, flow-sensitive type checking (e.g.\ for
linear types), and to improve the implementation of the lowering process.
Similarly, machine learning systems typically use ``ML graphs'' as a
domain-specific abstraction in the same way.

While the development of domain specific IRs is a well studied art, their
engineering and implementation cost remains high. The quality of the
infrastructure is not always a first priority (or easy to justify) for
implementers of these systems. 
Consequently, this can lead to lower quality compiler systems, including
user-visible problems like slow compile times, buggy implementations, suboptimal
diagnostic quality, poor debugging experience for optimized code, etc.

The \MLIR project aims to directly tackle these programming language design and
implementation challenges---by making it very cheap to define and introduce
new abstraction levels, and provide ``in the box'' infrastructure to solve
common compiler engineering problems. \MLIR does this by (1) standardizing the
Static Single Assignment (SSA)-based IR data structures,
(2) providing a declarative system for defining
\textit{IR dialects}, and (3) providing a wide range of common infrastructure
(including documentation, parsing and printing logic, location
tracking, multithreaded compilation support, pass management, etc).

This paper explores various design points of the \MLIR system, relates our
experience applying it to a number of different problems, and discusses
implications this work may have for language design and education.

\subsection{Contributions}

While most of the \MLIR system is built out of well known compiler algorithms,
the design points are sufficiently novel that it provides opportunities
for interesting research.  The contributions of this paper are:

\begin{itemize}
\item a description of a novel compiler infrastructure with important industrial and research applications;
\item new approaches to building scalable and modular compiler systems;
\item exploration of selected applications of \MLIR to diverse domains, illustrating the generality of the system;
\item shared experience developing systems that build on the \MLIR infrastructure.
\end{itemize}

\subsection{Where did \MLIR come from?}

\begin{figure}[h]
  \centering
	\includegraphics[width=0.7\textwidth]{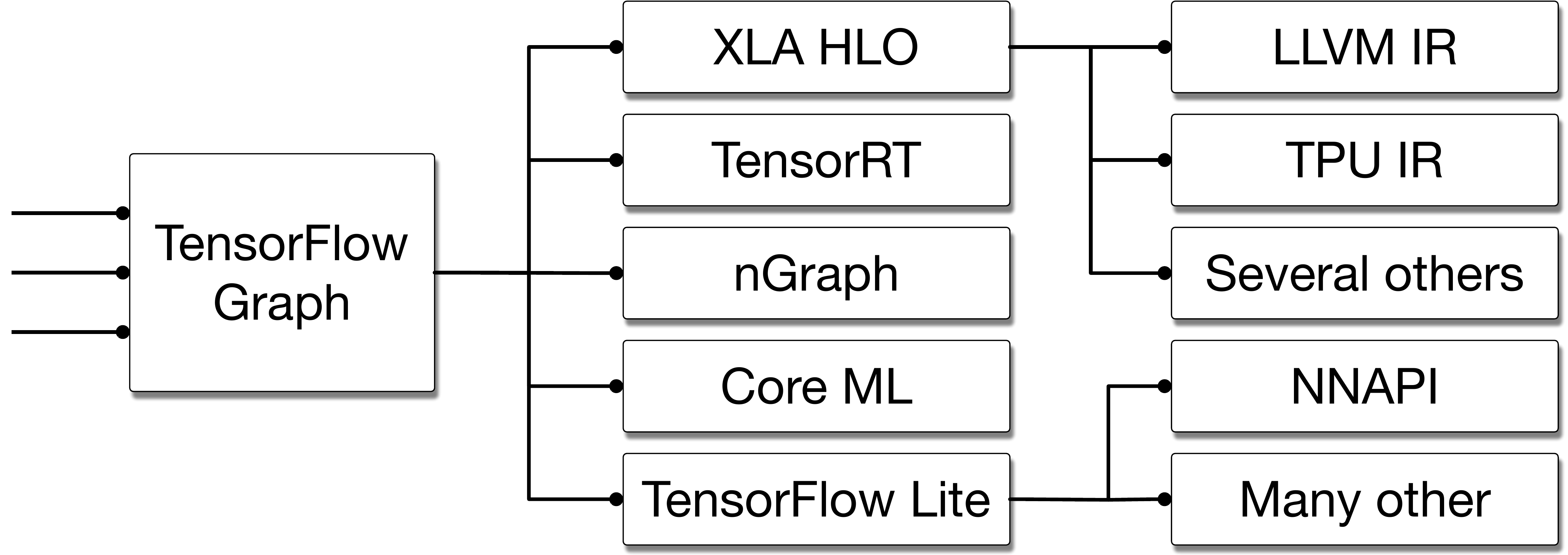}
	\caption{TensorFlow model execution spanning different frameworks.}
  \label{fig:ml-compile}
\end{figure}

Work on \MLIR began with a realization that modern machine learning frameworks
are composed of many different compilers, graph technologies, and runtime systems
(see~\figref{ml-compile})---which did not share a common infrastructure or design
point, and not all of which were following best practices in compiler design.
This manifested in multiple user-visible ways, including poor error messages,
failures in edge cases, unpredictable performance, and difficulty generalizing the
stack to support new hardware.

\begin{figure}[h]
  \centering
	\includegraphics[width=0.7\textwidth]{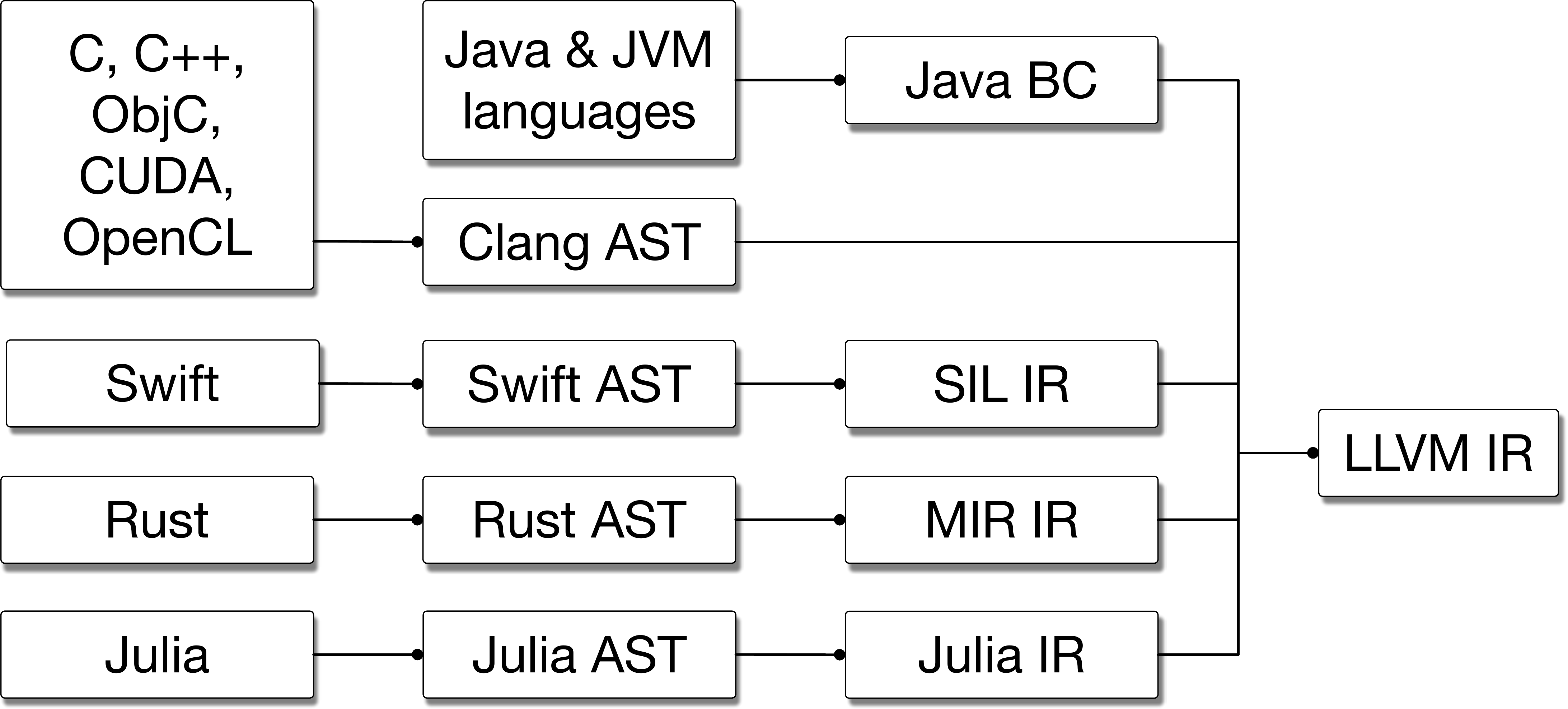}
	\caption{Compilation pipeline of different languages with multiple mid-level
    IRs for language-specific optimization with
    common backend for multiple hardware targets.}
   \label{fig:pl-midlevel}
\end{figure}

We soon realized that the compiler industry as a whole has a similar problem:
existing systems like LLVM are very successful at unifying and integrating work
across a range of different language implementations, but modern high level
languages often end up building their own high-level IR
and reinventing a lot of the same kinds of technology for higher levels of
abstraction (see~\figref{pl-midlevel}).  At the same time, the LLVM community
frequently struggled with questions about how to best represent parallel
constructs, how to share implementation of common front-end lowering infrastructure
(e.g.\ for C calling conventions, or cross-language features like OpenMP) with
no satisfactory solutions being available.

Faced with this challenge and perspective, we decided that we could not afford the
engineering effort to implement $N$ improved compiler instances, and we needed
to build a more general solution.  We reasoned that this would give us the ability
to invest in one set of high quality infrastructure which would benefit multiple
domains, would allow us to progressively upgrade existing systems in place,
would make it easier for us to tackle pressing problems like heterogeneous
compilation for specialized accelerators, and would provide interesting
research opportunities to define and explore.

Now that we have a significant amount of experience building and deploying
\MLIR-based systems, we are able to look back on the rationale and design of the infrastructure, and discuss why this direction was pursued.

\maybebreak
\section{Design Principles}\label{sec:design}


Let us now explore the requirements that guided the design of \MLIR.

\paragraph{Little builtin, everything customizable}

The system is based on a minimal number of fundamental concepts, leaving
most of the intermediate representation fully customizable. A handful of
abstractions---types, operations and attributes, which are the most common in
IRs---should be used to express everything else,
allowing fewer and more consistent abstractions that are easy to comprehend,
extend and adopt. Broadly, customizability ensures the system can adapt to
changing requirements and is more likely to be applicable to future problems.
In that sense, we ought to build an IR as a rich infrastructure with reusable components and programming abstractions supporting the syntax and semantics of its intermediate language.

A success criterion for customization is the possibility to express a diverse
set of abstractions including machine learning graphs, ASTs, mathematical
abstractions such as polyhedral, Control Flow Graphs (CFGs) and instruction-level IRs such as LLVM
IR, all without hard coding concepts from these abstractions into the system.

Certainly, customizability creates a risk of internal fragmentation due to
poorly compatible abstractions. While there is unlikely a purely technical
solution to the ecosystem fragmentation problem, the system should encourage
one to design reusable abstractions and assume they will be used outside of
their initial scope.

\paragraph{SSA and regions}

The Static Single Assignment (SSA) form \cite{Cytron:1991:ECS:115372.115320} is a
widely used representation
in compiler IRs.  It provides numerous advantages including making dataflow
analysis simple and sparse, is widely understood by the compiler community for its relation with continuation-passing style,
and is established in major frameworks.  While many existing IRs use a flat, linearized
CFG, representing higher level abstractions push introducing \emph{nested
regions} as a first-class concept in the IR.
This goes beyond the traditional region formation to lift higher level abstractions (e.g., loop trees), speeding up the compilation process or extracting instruction, or SIMD parallelism~\cite{Johnson:1994:PST:178243.178258,DBLP:conf/hpca/HavankiBC98,DBLP:journals/toplas/Ramalingam02}. To support heterogeneous compilation, the system has to support
the expression of structured control flow, concurrency constructs,
closures in source languages, and many other purposes. One specific challenge is to make CFG-based analyses and transformations compose over nested regions.

In doing so, we aim to sacrifice the normalization, and sometimes the canonicalization properties of LLVM. Being able to lower a variety of data and control structures into a smaller collection of normalized representations is key to keeping compiler complexity under control. The canonical loop structure with its pre-header, header, latch, body, is a prototypical case of a linearized control flow representation of a variety of loop constructs in front-end languages. We aim at offering users a choice: depending on the compilation algorithm of interest, of the pass in the compilation flow, nested loops may be captured as nested regions, or as linearized control flow. By offering such a choice, we depart from the normalization-only orientation of LLVM while retaining the ability to deal with higher level abstractions when it matters. In turn, leveraging such choices raises questions about how to control the normalization of abstractions, whith is the purpose of the next paragraph.

\paragraph{Progressive lowering}

The system should support \emph{progressive lowering}, i.e.\ from the
higher-level representation down to the lowest-level, with the lowering being performed
in small steps along multiple abstraction levels. The need for multiple levels
of abstractions stems from the variety of platforms and programming models that
a generic compiler infrastructure has to support.

Previous compilers have been introducing multiple fixed levels of abstraction in
their pipeline---e.g.\ the Open64 WHIRL representation \cite{open64} has five levels,
as does the Clang compiler which lowers from ASTs to LLVM IR,
to SelectionDAG, to MachineInstr, and to MCInst.  Whereas these approaches are
done in a rigid way, more flexible designs are required to support
extensibility.

This has deep implications on the
phase ordering of transformations. As compiler experts started implementing
more and more transformation passes, complex interactions between these passes
started appearing. It was shown early on that combining optimization passes
allows the compiler to discover more facts about the program. One of the first
illustrations of the benefits of combining passes was to mix constant
propagation, value numbering and unreachable code
elimination~\cite{Click1995}.
More generally, compiler passes can be roughly categorized into four roles:
(1) optimizing transformations, (2) enabling transformations, (3) lowering and
(4) cleanup. The system should allow for mixing and matching these roles at the
granularity of a single operation rather than sequencing passes on the full
compilation unit.

\paragraph{Maintain higher-level semantics}

The system needs to retain higher-level semantics and structure of computations
that are required for analysis or optimizing performance. Attempts to raise
semantics once lowered are fragile and shoehorning this information into a
low-level often invasive (e.g., all passes need to be verified/revisited in the
case of using debug information to record structure). Instead, the system should
maintain structure of computation and progressively lower to the hardware
abstraction. The loss of structure is then conscious and happens only where the
structure is no longer needed to match the underlying execution model.
For example, the system should preserve the structured control flow such as
loop structure throughout the relevant transformations; removing this
structure, i.e. going to CFG-based control flow, essentially means no further
transformations will be performed on this level.
The state of the art in modeling parallel computing constructs in a production compiler highlights how difficult the task may be in general~\cite{Khaldi:2015:LPI:2833157.2833158,Schardl:2017:TEF:3155284.3018758}.

As a corollary, mixing different levels of abstractions and different concepts in
the same IR is a key property of the system to allow a part of the
representation to remain in higher-level abstraction while another part is
lowered. This would enable, for instance, a compiler for a custom accelerator to reuse
some higher-level structure and abstractions defined by the system alongside
with primitive scalar/vector instructions specific to the accelerator.

\paragraph{IR validation}

The openness of the ecosystem calls for an extensive validation
mechanism. While verification and testing are useful to detect compiler bugs, and to capture IR
invariants, the need for robust validation methodologies and tools is amplified in an extensible system. The mechanism should aim to make this easy to define and as declarative
as practical, providing a single source of truth.

A long term goal would be to reproduce the successes of translation
validation
\cite{DBLP:conf/tacas/PnueliSS98,Necula:2000:TVO:358438.349314,DBLP:conf/popl/TristanL08,DBLP:conf/pldi/TristanL09}
and modern approaches to compiler testing
\cite{DBLP:conf/pldi/ChenGZWFER13}. Both are currently open problems in the context of an extensible
compiler ecosystem.

\paragraph{Declarative rewrite patterns}

Defining representation modifiers should be as simple as that of new
abstractions; a compiler infrastructure is only as good as the
transformations it supports. Common transformations should be implementable as
rewrite rules expressed declaratively, in a machine-analyzable format to reason
about properties of the rewrites such as complexity and completion. Rewriting systems have been studied extensively for their soundness and efficiency, and applied to numerous compilation problems, from type systems to instruction selection. Since we aim for unprecedented extensibility and incremental lowering capabilities, this opens numerous avenues for modeling program transformations as rewrite systems. It also raises interesting questions about how to represent the rewrite rules and strategies, and how to build machine descriptions capable of steering rewriting strategies through multiple levels of abstraction. The system needs to address these questions while preserving extensibility and enforcing a sound, monotonic and reproducible behavior.

\paragraph{Source location tracking and traceability}

The provenance of an operation---including its original location and applied
transformations---should be easily traceable within the system. This intends to
address the lack-of-transparency problem, common to complex compilation
systems, where it is virtually impossible to understand how the final
representation was constructed from the original one.

This is particularly problematic when compiling safety-critical and sensitive applications, where tracing lowering and optimization steps is an essential component of software certification procedures \cite{compcert}. When operating on secure code such as cryptographic protocols or algorithms operating on privacy-sensitive data, the compiler often faces seemingly redundant or cumbersome computations that embed a security or privacy property not fully captured by the functional semantics of the source program: this code may prevent the exposure of side channels or harden the code against cyber or fault attacks. Optimizations may alter or completely invalidate such protections \cite{Vu20}; this lack of transparency is known as WYSINWYX \cite{balakrishnan} in secure compilation. One indirect goal of accurately propagating high-level information to the lower levels is to help support secure and traceable compilation.

\maybebreak
\section{IR Design Details}\label{sec:structure}

\begin{figure}[t]
  \centering
		\includegraphics[width=.85\columnwidth]{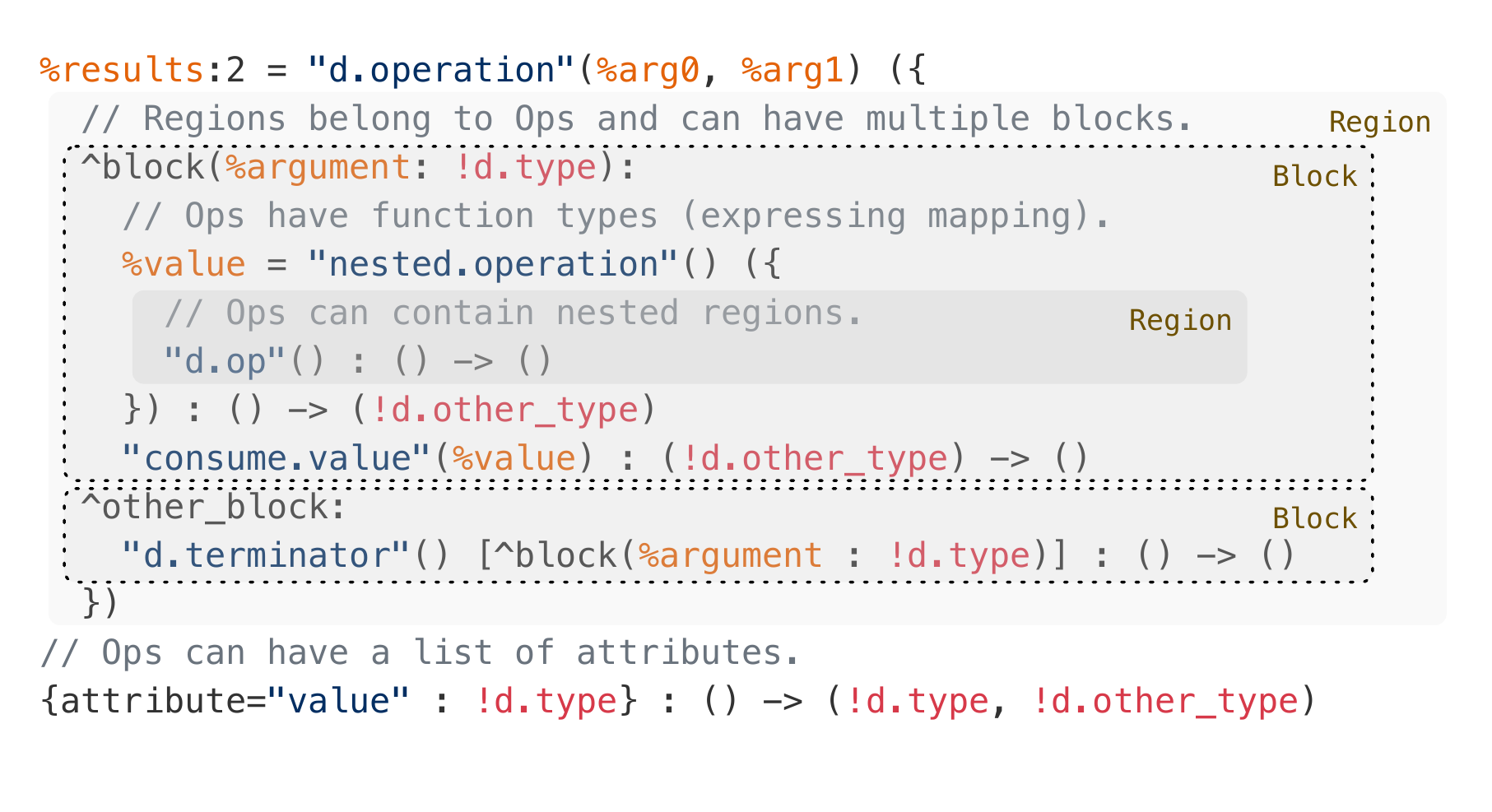}
  \caption{Operation (Op) is a main entity in \MLIR; operations contain a list
        of regions, regions contain a list of blocks, blocks contains a list of
        Ops, enabling recursive structures}
  \label{fig:structure}
\end{figure}

This section describes the design of the IR in \MLIR following the
principles from the previous section.

\begin{figure}[t]
  \centering
  \includegraphics[width=.8\textwidth]{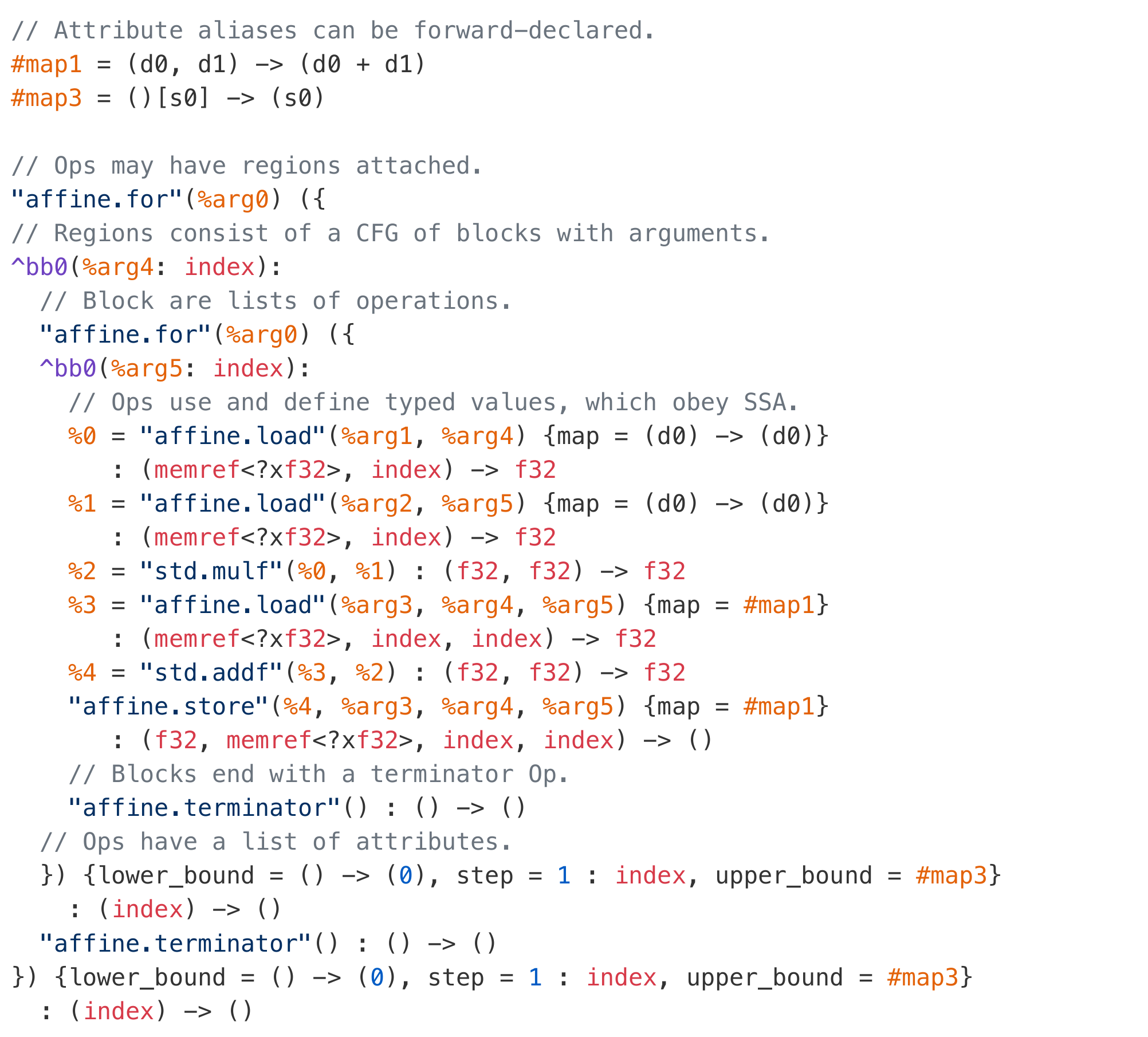}
  \caption{\MLIR \emph{generic} representation for polynomial multiplication
           using affine and std dialects. The same IR is displayed with the
           custom syntax~\figref{affine}.}
  \label{fig:generic}
\end{figure}

\paragraph{Operations}

The unit of semantics in \MLIR is an ``operation'', referred to as \emph{Op}.
Everything from ``instruction'' to ``function'' to ``module'' are modeled
as Ops in this system.  \MLIR does not have a fixed set of Ops, but
allows (and encourages) user-defined extensions---compiler passes
treat unknown Ops conservatively, and \MLIR has rich support for describing the
semantics of Ops to passes through traits, privileged operation hooks and
optimization interfaces as described in \secref{reusable_passes}.

Ops (see~\figref{structure}) have a unique opcode, which, textually, is a
string identifying its dialect and the operation.  Ops take and produce zero or more
\emph{values}, called \emph{operands} and \emph{results} respectively, and these
are maintained in SSA form. All values have a type, similarly to LLVM IR.
In addition to an opcode, operands and results, Ops
may also have \emph{Attributes}, \emph{Regions}, \emph{Block Arguments}, and
\emph{Location Information} as well. \figref{generic} illustrates values and Ops,
\lstinline|%|-identifiers are (packs of) named values, with ``\lstinline|:|''
specifying the number in a pack if more than one and ``\lstinline|#|'
a particular value. In the generic textual representation, operation names are
quoted string literals followed by operands in parentheses.

\paragraph{Attributes}

An \MLIR attribute is structured compile-time static
information, e.g., integer constant values, string data,
or a list of constant floating point values.  Attributes are typed, and each
Op instance has an open key-value dictionary from string names to attribute
values. In the generic syntax, attributes are between the Op
operands and its type as a brace-enclosed comma-separated list of key-value
pairs. For example, \figref{generic} uses attributes to define bounds
of a loop that are known to be constant affine forms:
\lstinline|{lower_bound = () -> (0), step = 1 : index,|
\lstinline| upper_bound = #map3}| where \lstinline|lower_bound|,
\lstinline|upper_bound| and \lstinline|step| are attribute names.
The \lstinline|() -> (0)| notation is used for inline affine forms, in this
case producing an affine function producing a constant \lstinline|0| value.
The \lstinline|#map3| notation is used for attribute \emph{aliases}, which
allow one to associate an attribute value with a label upfront and use the
label anywhere an attribute value is expected.

As with opcodes, there is no fixed set of attributes.  Attributes derive their
meaning either from the Op semantics or from the dialect (\secref{dialects}) they
are associated with.  Attributes are also extensible, allowing direct references
to foreign data structures, which is useful for integrating with existing systems.
For example, an attribute may refer to the contents of (known at compile time)
data storage in an ML system.

\paragraph{Location information}

\MLIR provides a compact representation for \emph{location information,}
and encourages the processing and propagation of this information
throughout the system.  It can be used to keep
the source program stack trace that produced an Op, to generate
debug information.  It standardizes the way to emit diagnostics from the compiler,
and is used by a wide range of testing tools.

Location information is also extensible, allowing a compiler to refer to
existing location tracking systems, high-level AST nodes, LLVM-style
file-line-column address, DWARF debugging info or whatever else is needed
for a high quality implementation.

\paragraph{Regions and blocks}

An instance of an Op may have a list of attached regions.  A \emph{region}
provides the mechanism for nested structure in \MLIR: a region contains a list
of blocks, and a block contains a list of operations (which may contain regions).
As with attributes, the semantics of a region are defined by the operation they
are attached to, however the blocks inside the region (if more than one) form a
Control Flow Graph (CFG).  For example, the \lstinline|affine.for| operation in
\figref{generic} is a loop with the single-block body attached as a region,
located between \lstinline|({| and \lstinline|})| delimiters.  The Op
specifies the flow of control across regions. In this example, the body is
executed repeatedly until the upper bound is reached.

The body of each region is a list of \emph{blocks}, and each block ends with a
\emph{terminator} operation, that may have \emph{successor} blocks to which the
control flow may be transferred. Each terminator
(e.g.\ ``switch'', ``conditional branch'' or ``unwind'') defines its
own semantics. It may chose to transfer the control flow to another block in
the same region, or return it to the Op enclosing the region. The graph of
successors defines a CFG, allowing standard SSA-based
control flow within a region.

Instead of using $\phi$ nodes, \MLIR uses a functional form of SSA~\cite{Appel98ssais}
where terminators pass values into \emph{block arguments} defined by the successor block.
Each block has a (potentially empty) list of typed block arguments, which are
regular values and obey SSA. The semantics of terminator Ops defines what
values the arguments of the block will take after the control is transferred.
For the first (entry) block of the region, the values are defined by the
semantics of the enclosing Op. For example, \lstinline|affine.for| uses the
entry block argument \lstinline|%arg4| as loop induction variable.

\paragraph{Value dominance and visibility}

Ops can only use values that are in scope, i.e. \emph{visible} according to
SSA dominance, nesting, and semantic restrictions imposed by enclosing
operations.  Values are visible within a CFG if they obey standard SSA
dominance relationships, where control is guaranteed to pass through a
definition before reaching a use.

Region-based visibility is defined based on simple nesting of regions:
if the operand to an Op is outside the current region, then it must be defined
lexically above and outside the region of the use.  This is what allows Ops
within an \lstinline|affine.for| operation to use values defined in outer scopes.

\MLIR also allows operations to be defined as \emph{isolated from above},
indicating that the operation
is a scope barrier---e.g.\ the ``std.func'' Op defines a function,
and it is not valid for operations within the function to refer to values
defined outside the function.  In addition to providing useful semantic
checking, a module containing isolated-from-above Ops may be processed
in parallel by an \MLIR compiler since no use-def chains may cross the
isolation barriers.  This is a important for compilation to utilize
multicore machines.

\paragraph{Symbols and symbol tables}

Ops can have a symbol table attached.  This table is a standardized way of
associating names, represented as strings, to IR objects, called \emph{symbols}. The
IR does not prescribe what symbols are used for, leaving it up to the Op
definition. Symbols are most useful for named entities need not obey SSA: they
cannot be redefined within the same table, but they can be used prior to their
definition. For example, global variables, functions or named modules can be
represented as symbols. Without this mechanism, it would have been impossible
to define, e.g., recursive function referring to themselves in their
definition. Symbol tables
can be nested if an Op with a symbol table attached has associated regions
containing similar Ops. \MLIR provides a mechanism to reference symbols from an
Op, including nested symbols.

\paragraph{Dialects}\label{sec:dialects}

\MLIR manages extensibility using \emph{Dialects}, which provide a logical
grouping of Ops, attributes and types under a unique namespace. Dialects
themselves do not introduce any new semantics but serve as a logical grouping
mechanism and can be used to provide dialect generic Op support (e.g.,
constant folding behavior for all ops in the dialect). The dialect namespace
appears as a dot-separated
prefix in the opcode, e.g., \figref{generic} uses \lstinline|affine| and
\lstinline|std| dialects.

The separation of Ops, types and attributes into dialects is conceptual
and is akin to designing a set of modular libraries. For example, a dialect can
contain Ops and types for operating on hardware vectors (e.g., shuffle,
insert/extract element, mask), and another dialect can contain Ops and types
for operating on algebraic vectors (e.g. absolute value, dot product, etc.).
Whether both dialects use the same vector type and where does this type belong
are design decisions left to the user of \MLIR.

While it is possible to put all Ops, types and attributes in a single dialect,
it would quickly become unmanageable due to the large number of simultaneously
present concepts and name conflicts, amongst other issues.
Although each Op, type and attribute belongs to exactly one dialect, \MLIR
explicitly supports a mix of dialects to enable progressive lowering. Ops from
different dialects can coexist at any level of the IR at any time, they can use
types defined in different dialects, etc. Intermixing of dialects allows for
greater reuse, extensibility and provides flexibility that otherwise would
require developers to resort to all kinds of non-composable workarounds.

\paragraph{Type system}

Each value in \MLIR has a type, which is specified in the Op
that produces the value or in the block that defines the value as an argument.
Types provide compile-time semantics for the IR. The type system in \MLIR is
user-extensible, and may refer to existing foreign type systems (e.g. an
\lstinline|llvm::Type| or a \lstinline|clang::Type|). \MLIR enforces strict type
equality checking and does not provide type conversion rules. Ops list their
inputs and result types using trailing function-like syntax. In \figref{generic},
\lstinline|std.load| maps from the memory reference and
index types to the type of the value it loads.

From the type theory point of view, \MLIR only supports non-dependent types,
including trivial, parametric, function, sum and product types. While it is
possible to implement a dependent type system by combining Ops with
symbols and user-defined types in a literal interpretation of Curry-Howard
isomorphism, such types will be opaque to the IR.

\paragraph{Standard types}

In addition, \MLIR provides a standardized set of commonly used types,
including arbitrary precision integers, standard floating point types, and simple
common containers---tuples, multi-dimensional vectors, and tensors.  These types
are merely a convenience that are useful to authors of dialects, but their use is not
required.

\paragraph{Functions and modules}

Similarly to conventional IRs, \MLIR is usually structured into functions and
modules.  However, these are not new or separate concepts in \MLIR: they are
implemented as Ops in the builtin dialect.

A module is an Op with a single region containing a single block, and
terminated by a dummy Op that does not transfer the control flow. A module defines a
symbol and can be referenced. Like any block, its body contains a list of Ops,
which may be functions, global variables, compiler metadata, or other top-level
constructs.

A function is an Op with a single region, with arguments corresponding to
function arguments. It defines a symbol and can be referenced by name. The
control flow is transferred into a function using a function call Op. Once
inside, the control flow follows the CFG of the blocks in the region. A
``return'' terminator does not have successors and instead terminates the
region execution, transferring the control flow back to the call-site of the
function.
Any operands of the ``return'' terminator Op are the returned values of the
function.

\maybebreak
\section{IR Infrastructure}\label{sec:infra}

Beyond the IR itself, \MLIR provides
infrastructure for defining IR elements such as dialects, Ops,
pattern rewrites, verification and reusable
passes. The infrastructure aspect of \MLIR is essential for providing
extensibility and ease of use when defining new abstractions and using
\MLIR as an optimization toolkit.

\subsection{Operation description}

\MLIR uses
TableGen-based~\cite{tablegen} specification for Operation Descriptions (ODS),
defining the structure of an Op and components of its verifier declaratively.
TableGen is a data modeling tool intended to help define and maintain
records of domain-specific information, used extensively in LLVM. We chose it
for modeling Ops and rewrite patterns to leverage its acceptance by the
industry. ODS can be seen as a DSL for \MLIR Op definition \emph{embedded} into
the TableGen input language, so the ODS syntax is imposed by TableGen, but the
\MLIR-specific semantics is provided by ODS. The ODS definition is ultimately
translated into \Cpp code (including Op classes with named accessors,
verification, etc.) which interoperate with the rest of the system.

\begin{figure}[t]
  \centering
  \includegraphics[width=.8\textwidth]{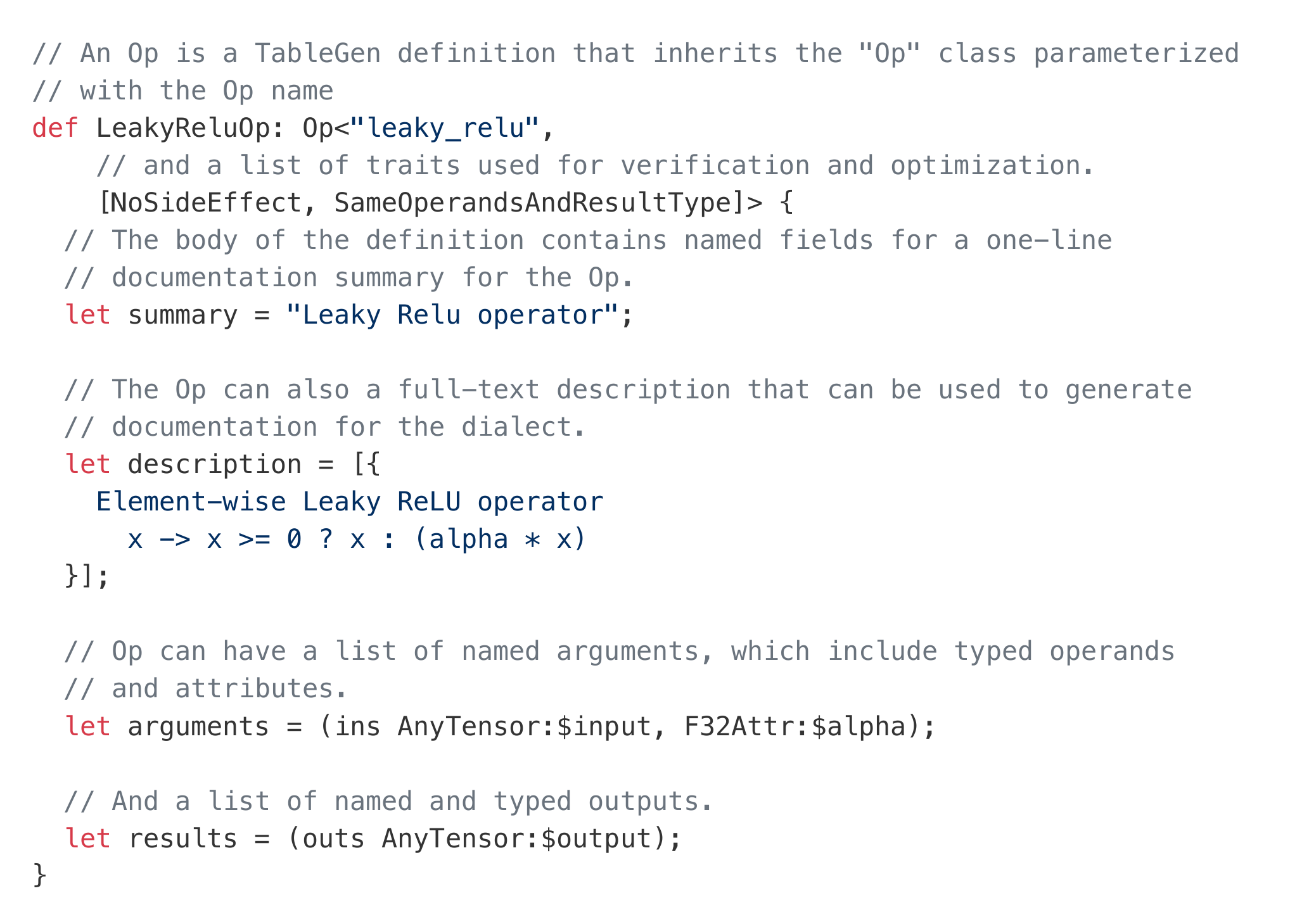}
  \caption{Operation Definition Syntax (ODS) provides a concise way of defining
        new Ops in \MLIR. Here, one defines the \lstinline|LeakyRelu| Op taking
        a tensor and a floating-point value, and returning a tensor of the same
        type as the input one.}
  \label{fig:ods}
\end{figure}

Ops are modeled in ODS using the  TableGen \lstinline|Op| class.
Figure~\ref{fig:ods} shows an example of Op ODS definition.
Each defined Op has a \emph{name} which is
a unique identifier, a list of \emph{traits} that describe Op properties, a list of
\emph{arguments} that specify Op's operands and attributes, and a list of Op's \emph{results}.
Arguments and results have names and type constraints (e.g., fixed-shape tensor of
\lstinline|float| or \lstinline|int32|). Op definition may also specify
human readable Op description for the documentation.
And a (limited) custom textual form for which custom printer/parser will be generated.
When Op definition
requires finer-grain control than ODS provides, additional \Cpp code can be
injected via \emph{builder}, \emph{printer},
\emph{parser}, \emph{verifier} clauses.
Op traits can be generic, e.g., ``has no side-effects'', and
dialect- or ODS-specific, e.g., ``has custom exporter''.
Traits in ODS may be
backed by \Cpp classes defining the behavior of the trait. There is no fixed
set of traits, but some traits are known by ODS (e.g., ``shape result and
operand type'' represents a constraint that fully captures the output type given
input types) or optimizers (e.g., ``has no side-effects'',
see~\secref{interfaces}).

Type constraints check properties of the type of arguments/results and are
user/dialect extensible. \MLIR infrastructure also provides numerous pre-defined
type constraints, such as ``any type'', ``tensor with element satisfying the
given constraint'', ``vector of given rank'', etc. ODS also has limited support
for automatically deducing the return type of results of operands using the
constraints induced by the traits, see~\secref{drr} for more information.

\subsection{Declarative rewrites}\label{sec:drr}

Many \MLIR transformations involve Op manipulations, and while some
transformations require complex modifications of the IR, many others can be
expressed as simple rewrites on the DAG defined by SSA use-def relations.  \MLIR provides a graph
rewriting framework complemented with the Declarative Rewrite Rule (DRR) system
that makes it simple to express patterns.

\begin{figure}[t]
  \centering
  \includegraphics[width=.8\textwidth]{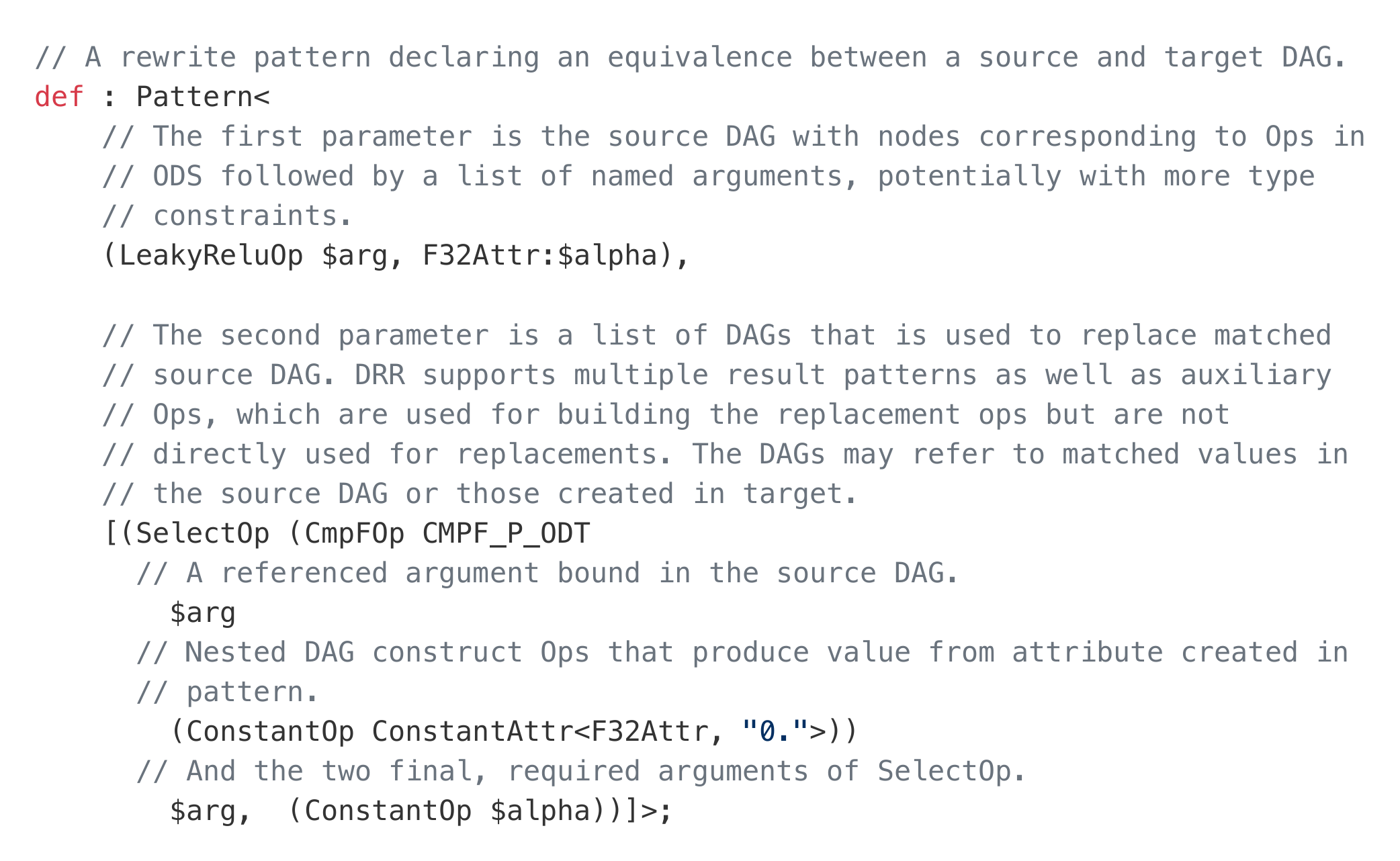}
  \caption{Declarative graph rewrite rule transforming a LeakyRelu into a
    Compare-Float, followed by a Select.}
  \label{fig:rewrite}
\end{figure}

Similarly to ODS, DRR is a DSL embedded into the TableGen language. DRR
expresses source and target DAG patterns along with constraints (including
dynamic constraints~\cite{Thier:2018:FFI:3178372.3179501}) and
benefits for pattern prioritization. Patterns can capture and reuse arguments
of an Op.  Conceptually, DRR expresses equivalence of DAGs under specific
constraints. \figref{rewrite} gives an example of a DRR pattern that converts
an Op defined in \figref{ods} into common lower-level implementation consisting
of a \lstinline|compare| and a \lstinline|select|.

DRR is converted into \Cpp code, which can be intermixed with more complex
patterns defined directly in \Cpp using the generic graph rewriting framework. This
ability allows \MLIR to keep the common case simple without restricting
generality of the framework.

\subsection{Pass manager}

The \MLIR pass manager organizes and handles the efficient execution of a series
of \emph{IR passes} operating on various granularities.  Whereas pass management
in existing systems is typically defined over a fixed granularity (e.g., module,
function or loop pass managers), in \MLIR modules and functions are not special---they are merely Ops with regions and there can be
multiple variants of them. Therefore, the \MLIR pass manager is also not
specialized on a fixed set of ops, but instead works on arbitrary Ops at
arbitrary levels of nesting.

\paragraph{Parallel compilation}\label{sec:parallel_comp}

An important requirement of \MLIR is the need to utilize multi-core
machines for faster compilation.  The pass manager supports
concurrent traversal and modification of the intermediate representation,
which is made possible by invariants provided by ``isolated-from-above''
property of operations, because the SSA use-def chains cannot cross the
region boundaries of these ops.  Operations with this behavior (e.g. the
``std.func'' operation) thus define a tree of regions that may be
processed in parallel.

This requirement is the reason why (in contrast to, for example, LLVM),
\MLIR does not feature whole-module use-def chains.  Global objects
are referenced through symbol table entries, and constants are implemented
as operations with associated attributes.

\subsection{Round-trippable textual IR form}\label{sec:roundtrippable}

\begin{figure}[t]
  \centering
  \includegraphics[width=.8\textwidth]{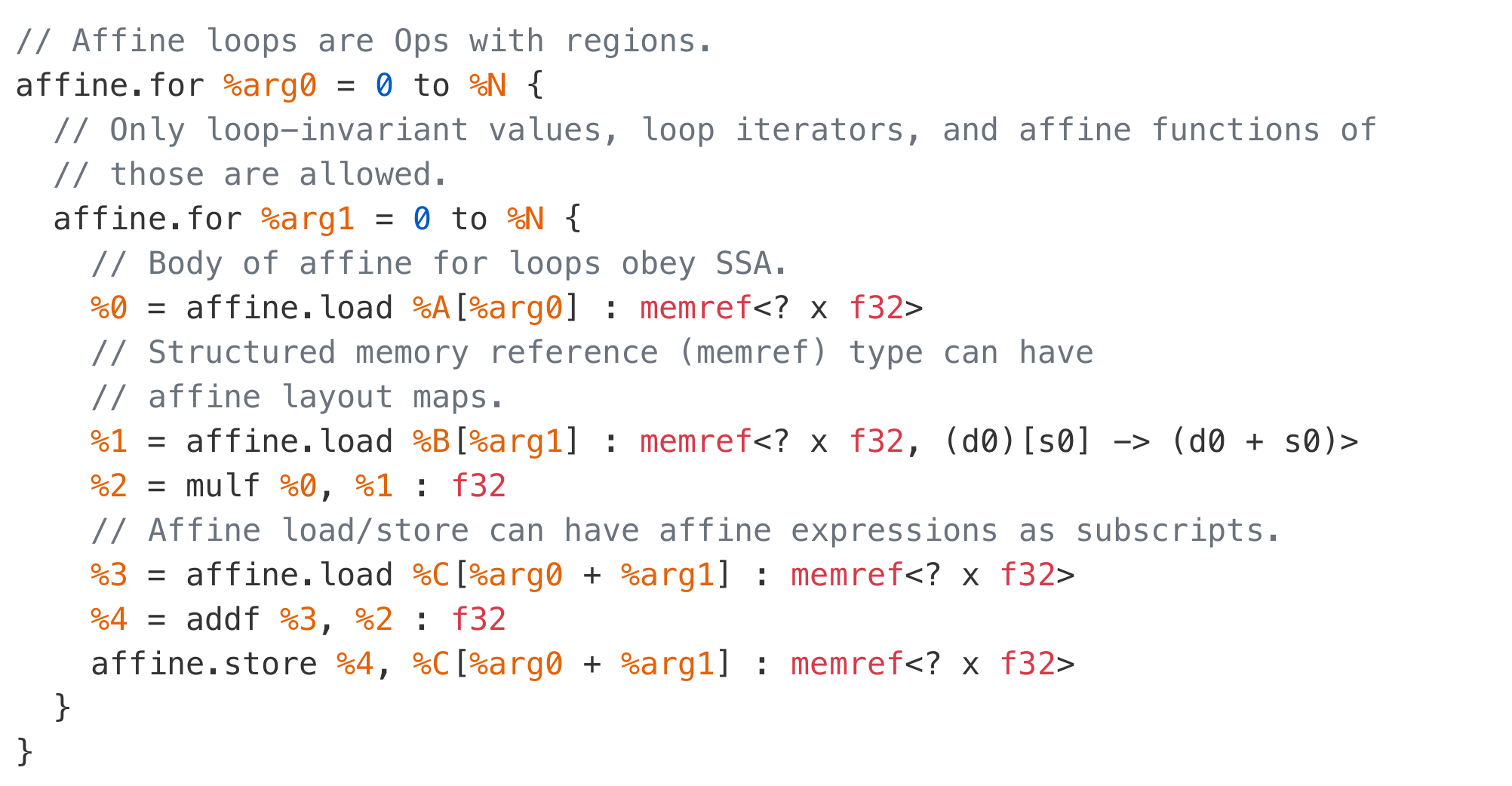}
  \caption{Representing polynomial multiplication kernel
    \lstinline|C(i+j) += A(i) * B(j)| using \MLIR affine dialect.}
  \label{fig:affine}
\end{figure}

The IR and Ops in \MLIR have a textual representation that fully reflects
the in-memory representation, which is critical for debugging, understanding
the IR during transformations, and for writing test cases.  The
raw IR form shown in \figref{generic} is verbose and difficult to understand.
Therefore \MLIR allows for defining custom printing and parsing formats for Ops,
which allows the example to be printed and parsed as shown in \figref{affine},
which is much easier to work with.

Both forms are fully round-trippable and each compiler pass may be tested
separately, using the textual form as both input and output. Since there is
no hidden state, the
result of running an individual pass is identical to running the same pass in
the full pass pipeline. This approach is user-friendly, as the IR form can be
created by hand, and IR transformations are easy to trace.

\subsection{Documentation}
Dialects, Ops and Interfaces have a documentation generated from their ODS
descriptions. Beyond one line \emph{summary} and more readable
\emph{description}, the generated documentation includes argument and result
type constraints. As the same source is used for both the verification code and
documentation, the documentation has higher changes of remaining in sync with
runtime behavior.

\subsection{Verifiers}

\emph{Verifiers} are used to enforce structural correctness
of the IR and invariants of Ops, allowing passes to assume a verified IR with
invariants checked, and also serves as a debugging tool. Verification
starts with checking structural properties of \MLIR overall: types must match
exactly, values are defined only once and obey dominance and visibility, symbol
names are unique in the symbol table, all blocks end with
terminator Ops, etc. After that, individual Op and attribute verifiers are applied.
Each Op may define a set of structural and semantic rules that check its
validity. For example, a binary Op checks that it has two operands, many Ops
only accept values of specific types, and many require specific
attributes or regions attached. Similarly, dialect attributes can be only
allowed on specific Ops or impose further restrictions on the Ops to which they
are attached. For example, a dialect attribute can require an Op to only use
types defined in the dialect even though the Op itself is more generic.
Verification failures are considered invariant violations and abort compilation.

\maybebreak
\section{Evaluation: Applications of \MLIR}

\MLIR is a system that aims to generalize and drive a wide range of
compiler projects, so our primary evaluation metric is to show that it
is being adopted and used for diverse projects.  We provide a summary
of community activity and describe a few use cases in more detail to
highlight the generality and extensibility of \MLIR and demonstrate
how well it implements the customizability design principle.

Today, \MLIR is a growing open source project with a community
spanning academia and industry.  For example, the academic workshop
about use of \MLIR in High-Performance Computing (HPC) was attended by
individuals from 16 universities and involved 4 national laboratories
from 4 different countries.  \MLIR was also endorsed by 14
multinational companies and at the LLVM Developer Meeting more than
100 industry developers attended a roundtable event about \MLIR.
Community adoption and participation is a proxy measure for usability
and need.  More than 26 dialects are in development in public or
private and 7 projects across different companies are replacing custom
infrastructure with \MLIR.  We argue that this shows a real need for
\MLIR, as well as endorses its usability.

\subsection{TensorFlow graphs}

While the other discussed representations are familiar to most compiler developments,
one of key use cases for \MLIR is to support the development of machine
learning frameworks. Their internal representations is often based on a data
flow graph~\cite{veen} with a dynamic execution semantics.

TensorFlow~\cite{tensorflow2015-whitepaper} is an example of such framework.
Its representation is a high-level dataflow computation where the
nodes are computations which can be placed on various devices, including
specific hardware accelerators.

\begin{figure}[t]
  \centering
  \includegraphics[width=.8\textwidth]{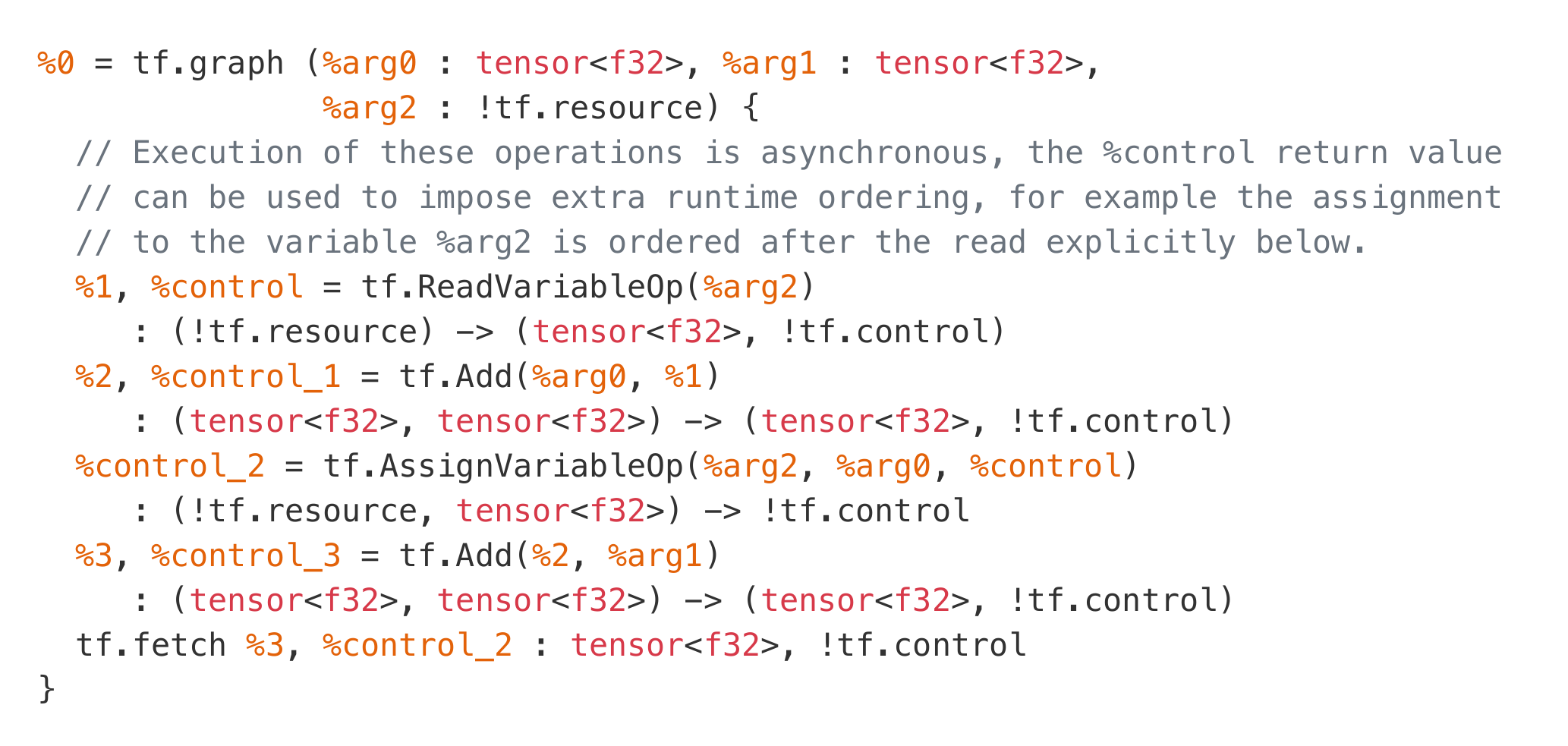}
%
\caption{SSA representation of a TensorFlow dataflow graph in \MLIR.}\label{fig:tf}
\end{figure}

\MLIR is used in TensorFlow to model this internal representation and perform
transformations for the use cases presented in \figref{ml-compile}: from
simple algebraic optimizations to retargeting graphs
for parallel execution on data center clusters of hardware accelerators, from
lowering to a representation suitable for mobile deployment to
generating efficient native code using tools like XLA~\cite{xla}. The representation of
a TensorFlow graph in \MLIR is illustrated on~\figref{tf}.

\subsection{Polyhedral code generation}\label{sec:affine}

One of the original motivations for \MLIR was the exploration of
polyhedral code generation for accelerators.
The affine dialect is a simplified polyhedral representation that was
designed to enable progressive lowering.
While a full exploration of the design points here is out of scope for this
paper, we illustrate aspects of the affine dialect to show the modeling power
of \MLIR and contrast the affine dialect with past
representations~\cite{Fea92b,uruk,isl,ppcg,tc}.

\subsubsection{Similarities}
The \MLIR affine dialect operates on a structured multi-dimensional
type for all accesses to memory.  In the default case, these structured types
are \emph{injective}: different indexings are guaranteed not to alias by
construction, a common precondition for polyhedral dependence analyses.

Affine modeling is split in two parts. Attributes are used to model affine
maps and integer sets at compile-time and Ops are used to apply affine
restrictions to the code. Namely, \lstinline|affine.for| Op is a ``for'' loop
with bounds expressed as affine maps of values required to be invariant in a
function. Thus loops have static control flow. Similarly,
\lstinline|affine.if| is a conditional restricted by affine integer sets.  The
bodies of loops and conditionals are regions that use \lstinline|affine.load|
and \lstinline|affine.store| to restrict indexing to affine forms of
surrounding loop iterators. This enables exact affine dependence analysis while
avoiding the need to infer affine forms from a lossy lower-level representation.

\subsubsection{Differences}
The differences with existing polyhedral frameworks are numerous, we can
characterize them in four categories:

(1) \emph{Rich types}: the \MLIR structured memory reference type contains a layout
map connecting the index space of the buffer to the actual address space.
This separation of concerns makes loop and data transformations compose better:
changes to data layout do not affect the code and do not pollute dependence
analysis. Such mixes of transformations have been explored
previously~\cite{chandan} but are uncommon.

(2) \emph{Mix of abstractions}: Bodies of affine loops in \MLIR can be expressed with
operations on typed SSA values.
Therefore, all traditional compiler analyses and transformations
remain applicable and can be interleaved with polyhedral transformations.  On
the contrary, polyhedral compilers often abstract such details away completely,
making it challenging for a polyhedral compiler to manipulate, e.g., vector
types.

(3) \emph{Smaller representation gap}: One of the key features of the polyhedral
model is its ability to represent the
order of loop iterations in the \emph{type system}. In this system, a large
number of loop transformations compose directly and can be reasoned about using
simple mathematical abstractions~\cite{uruk}. However, polyhedral
transformations require \emph{raising} into a representation often
drastically different from the original~\cite{polly,pollytactics}. Furthermore,
the conversion from
transformed polyhedra to loops is computationally hard~\cite{cloog}.
\MLIR-based representation maintains high-level loop structure around
lower-level representation, removing the need for raising.

(4) Compilation speed is a crucial goal for \MLIR as discussed in
\secref{parallel_comp}, but has not been a focus of most existing polyhedral
approaches.  These rely heavily on algorithms with exponential complexity: on integer
linear programming to derive loop orderings automatically and on polyhedron
scanning algorithms to convert the representation back to loops.
The approach taken by \MLIR explicitly does not rely on polyhedron scanning
since loops are preserved in the IR.

Experience with the affine dialect shows that it is useful for a wide range
of code generation projects, and its development was important exploration
that the \MLIR design made practical.

\subsection{Fortran IR (FIR)}

The LLVM Fortran frontend ``flang'' is currently under major development,
led by NVIDIA/PGI. Similar to Swift, Rust, and others, flang needs
a specialized IR in order to support advanced transformations
for high-performance Fortran codebase, and is using \MLIR to support
these Fortran-specific optimizations~\cite{flang}.  These high-level
optimizations---advanced loop
optimizations, array copy elimination, call specialization,
devirtualization---would be hard implement using only LLVM.

\begin{figure}[h]
  \centering
  \includegraphics[width=.7\textwidth]{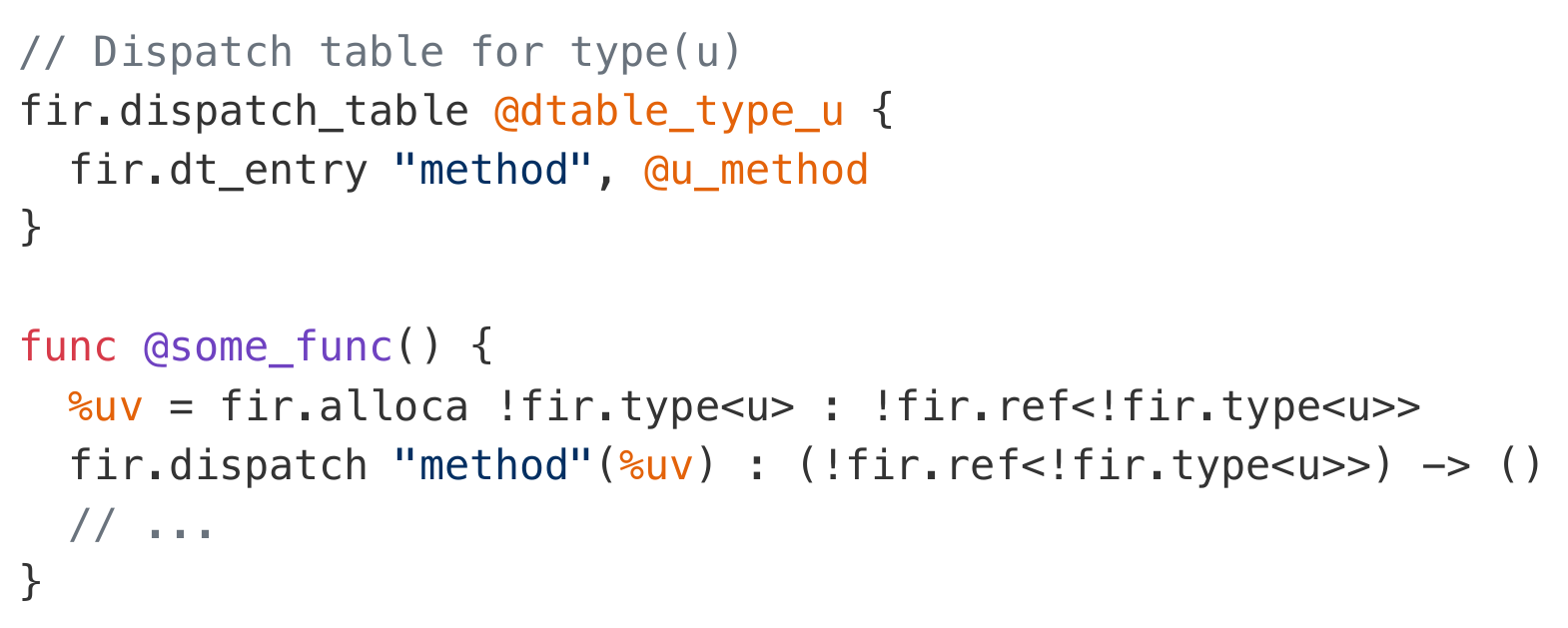}
%

\caption{FIR has first class support for dynamic virtual function
dispatch table.}\label{fig:fir}
\end{figure}

For example, FIR is able to model Fortran virtual dispatch table as a first
class concept (see~\figref{fir}).

The ability to model the high-level semantics of the programming language in a
structured IR is very powerful. For example, first-class modeling of the
dispatch tables allows a robust devirtualization pass to be implemented.  While
this could have been implemented with a bespoke compiler IR, the
use of \MLIR allowed the flang developers to spend their engineering resources
focused on the IR design for their domain instead of reimplementing basic
infrastructure.

The choice of \MLIR also unlocks the reusability of other dialects that are not
specific to Fortran: a language-independent OpenMP dialect could be shared
between Fortran and C language frontends.  Similarly, targeting a
heterogeneous platform using OpenACC
becomes tractable within \MLIR through the sharing and reuse of the
GPU-oriented dialects and passes. This is straightforward thanks to \MLIR 
begin specifically designed to support a mix of composable dialects.

\subsection{Domain-specific compilers}

The applications of \MLIR above are within large compilation
workflows, but it is also used to build domain specific compilers for
specific small workflows.  A reusable and modular infrastructure makes
these specialized paths feasible and relatively cheap to build.

\paragraph{Optimizing \MLIR pattern rewriting}

\MLIR has an extensible system for pattern rewrites described in~\secref{infra}.
In addition to statically declared patterns, we had
applications where the rewrite patterns needed to be dynamically
extensible at runtime, allowing hardware vendors to add new lowerings
in drivers.  The solution was to express \MLIR pattern rewrites as an
\MLIR dialect itself, allowing us to use \MLIR infrastructure to build
and optimize efficient Finite State Machine (FSM) matcher and
rewriters on the fly.  This work includes FSM optimizations seen in
other systems, such as the LLVM SelectionDAG and GlobalISel
instruction selection systems.


\paragraph{Lattice regression compiler}

Lattice regression~\cite{NIPS2009_3694} is a machine learning
technique renowned for fast evaluation times and interpretability.
The predecessor of the compiler was implemented using \Cpp
templates. This allowed for high-performance code with
metaprogramming, but expressing general optimizations on the
end-to-end models was not straightforward. This particular lattice
regression system is used in applications with multiple millions of
users and hence performance improvements are critical.

\MLIR was used as the basis for a new compiler for this specialized
area, which was driven by a specialized search approach---effectively
resulting in a machine learning problem being solved during
compilation. The resultant compiler was developed by investing a  3
person-month effort, and resulted in up to $8\times$ performance improvement
on a production model, while also improving transparency during
compilation.

\maybebreak
\section{Consequences of the \MLIR Design}

The \MLIR design facilitates the modeling of new language and compilation abstractions while reusing existing, generic ones as well as their associated compilation methods. Effectively,
the solution to many problems is to ``add new ops, new types'', possibly collected into
``a new dialect''.  This is a significant design shift for compiler engineering.
It produces new opportunities, challenges, and insights.  This section
explores a few of them.

\subsection{Reusable compiler passes}
\label{sec:reusable_passes}

The ability to represent multiple levels of abstraction in one IR creates the
natural desire to write passes that work across multiple levels of abstraction. A
common question about \MLIR is ``how do you write a compiler pass when you have
openly extensible operations and type system?'' While it is always possible
for a compiler pass to treat unknown constructs in a conservatively correct
way, our goal is to produce high performance code, so we need to do
useful things in common cases. We have found four major approaches:

\paragraph{Fundamental operation traits}

Some ``bread and butter'' compiler passes like Dead Code Elimination and Common
Subexpression Elimination rely on very simple properties (like ``has no
side effect'' or ``is commutative'') that we define as Op traits.  The definition
of an operation in ODS allows the author of the operation to specify these
traits, and passes can uses this information to remain applicable
across many different abstraction domains.

\MLIR is extensible enough that it has a few structural properties, including
information about whether an operation is known to be a control flow terminator,
whether an operation containing a region is known to be isolated-from-above, etc.
These allow generic modeling and
processing of functions, closures, modules, and other code structures.

\paragraph{Privileged operation hooks}

While certain traits can be modeled with a single bit, others need \Cpp code to
provide an implementation---constant folding logic for example. \MLIR has first
class support for certain hooks applicable to a large number of passes. These
hooks can either be implemented on a per-operation basis, or in
the Dialect object itself. The later approach is
convenient for things like constant folding of TensorFlow ops, where delegation
to existing logic is straightforward.

While constant folding is very important functionality, a more interesting hook
is \lstinline|getCanonicalizationPatterns|, which allows one to specify folding
patterns that apply to an operation.  This enables open extensibility of important
algebraic simplifications (e.g. $x-x \rightarrow 0$, $\min(x,y,y) \rightarrow \min(x,y)$ etc.) and powers
a common ``Canonicalization'' pass that can now be applied to all dialects.  This
allows this single extensible system to subsume things like
``InstCombine'', ``DAGCombine'', ``PeepholeOptimizer'', ``SILCombine'', and
other special purpose passes seen in the LLVM ecosystem (and other
compilers), which are a well-known maintenance and complexity burden.

\paragraph{Optimization interfaces}\label{sec:interfaces}

A primary goal of \MLIR is to allow open extensibility---not just in terms of
operations and types, but also in terms of transformations.  While canonicalization
and constant folding are critical operations, there are a number of standard
transformations that need to be parameterized in certain ways---e.g., to describe
transformation specific properties, to implement cost models, etc.

The solution is a subsystem known as ``Optimization Interfaces''. Consider the
\MLIR inlining pass: we would like the inliner to work on TensorFlow graphs,
Flang functions, closures in a functional language etc.---but the inliner does
not know what call sites or even the callees are! The core characteristics that
an inliner needs to know are:

\begin{itemize}
\item whether it is valid to inline an operation into a given region;
\item how to handle terminator operations that ended up in the middle of a block
after inlining.
\end{itemize}

\begin{figure}[h]
  \centering
  \includegraphics[width=.85\textwidth]{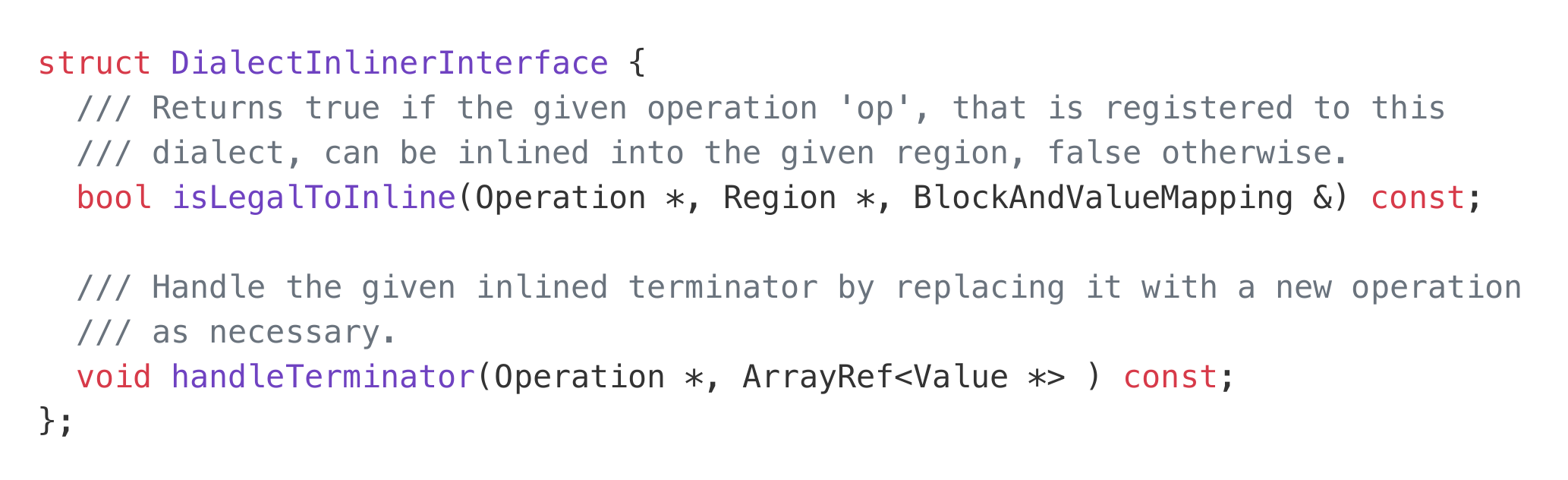}
	%
\caption{Dialect interface to query legality of inlining.}\label{fig:inlining}
\end{figure}

In order to know these properties, the Inliner pass defines the interface
in~\figref{inlining}. Individual operations and dialects may register their
implementation of this interface with \MLIR to benefit from the generic inliner
pass. If an operation or dialect fails to provide an interface then the
corresponding optimization pass will treat the operation conservatively. This
design allows the implementer of a dialect to get up and
running quickly, but derive more value out of the system by putting more
implementation effort into interfaces likes these over time.

Optimization interfaces also provide modularity benefit for the core compiler, because the
dialect specific logic is implemented within the dialects themselves, instead of
inside the core transformations.

\paragraph{Dialect specific passes}

Finally, it is valid and useful to define passes that are specific to particular
dialects, which can be driven by full semantics of operations in the dialect(s) they are
designed for.  These passes are just as useful in the \MLIR system as they
are in other compiler systems.  For example, code generators that want to do
custom scheduling of machine instructions based on particular machine constraints
or other tricks that do not fit into a broader framework. This is a simple and
useful starting point for new transformations, where generalization isn't required.

\subsection{Mixing dialects together}

One of the most profound (but also most difficult to grok) aspects of \MLIR
is that it allows and encourages mixing operations from different dialects
together into a single program.  While certain cases of this are reasonably
easy to understand (e.g. holding host and accelerator computation in the same
module) the most interesting cases occur when dialects are directly mixed---
because this enables an entire class of reuse that we have not seen in other
systems.

Consider the affine dialect described in \secref{affine}. The definition
of affine control flow and affine mappings are independent of the semantics of
the operations that are contained in affine regions.  In our case, we combine
the affine dialect with the ``standard'' dialect that represents simple arithmetic
in a target independent form like LLVM IR, with multiple target-specific machine
instruction dialects for internal accelerators. Others have combined it with
abstractions from other problem domains.

The ability to reuse generic polyhedral transformations (using Op interfaces to
get semantics of operations in specific transformations) is a powerful (and
exciting to us) way of factoring compiler infrastructure. Another
example is that an OpenMP dialect could be used and reused across a wide variety
of source-language IRs.

\subsection{Interoperability}

Our work involves interoperation with a large number of existing systems, e.g.,
machine learning graphs encoded as protocol buffers, compiler IRs including
LLVM IR, proprietary instruction sets, etc.  Often the representation has a
number of suboptimal or unfortunate decisions that made sense in the context of
an existing system, but capabilities of \MLIR enable a more expressive
representation. Because importers and exporters are notoriously
difficult to test (often the test cases are binary formats), we want to make
sure their complexity is minimized.

The solution is to define a dialect that corresponds to the foreign system as
directly as possible---allowing round tripping to-and-from that format in a simple
and predictable way. Once the IR is imported into \MLIR form, it can be raised
and lowered to a more convenient IR using all of the \MLIR infrastructure for doing
these transformations, and allows those transformations to be tested similarly
to all the other \MLIR passes are.

There are numerous examples of such dialects, including: a) the LLVM dialect---which
maps LLVM IR into \MLIR, b) the representation of TensorFlow graphs---which
is raised to ease analysis and transformations related to ``switch and merge''
nodes in TensorFlow, and c) functional-style control flow operators---``functional
while'' and ``functional if'' are common in machine learning graphs, in which it is
more convenient to work with their bodies as regions instead of out-of-line
functions.

This approach has worked well for us, and the \MLIR tooling has also been useful
to write tests for these foreign binary file formats.

\subsection{Unopinionated design provides new challenges}

While \MLIR allows one to define almost arbitrary abstractions, it provides very
little guidance on what \emph{should} be done: what works better or worse in
practice? We now have some experience with a number of engineers and researchers
applying the techniques and technologies to new problem domains, and have
realized that the ``art'' of compiler IR design and abstraction design is not
well understood in the compiler and languages field---many people work within
the constraints of established systems, but relatively few have had the
opportunity define the abstractions themselves.

This is a challenge, but is also another set of opportunities for future research.
The broader \MLIR community is building a significant amount of expertise with these
abstraction design trade-offs, and we expect this to be a fertile area of study over
time.

\subsection{Looking forward}

The design of \MLIR is different enough from other compiler infrastructures that
we are still learning---even after building and applying it to many different
systems. We believe that there is still a lot to discover, and several more
years of research will be required until the design points are all fully
understood and best practices are established. For example, the rise of
out-of-tree dialects, increasing number of source language frontends using
\MLIR, possible application to Abstract Syntax Trees, and applications to
structured data (like JSON, protocol buffers, etc) which are still very early
and are likely to uncover interesting new challenges and opportunities.

\maybebreak
\section{Related Work}

\MLIR is a project that overlaps with multiple different domains. While the
composed infrastructure provides a novel system, individual components have analogs in the literature. For references and discussion directly related to the IR design itself, please refer to Section~\ref{sec:design}.

\MLIR is a compiler infrastructure akin to
LLVM~\cite{LLVM:CGO04}, but where LLVM has been a great boon to scalar
optimizations and homogeneous compilation, \MLIR aims to model a rich set of data structures and algorithms as first-class values and operations, including tensor algebra and algorithms, graph representations, as well as heterogeneous compilation.
\MLIR allows mix-and-match optimization decomposing compilation passes into components and redefining lowering, cleanup roles. This is largely attributed to the pattern rewriting infrastructure, capturing full-fledged transformations as a composition of small local patterns and controlling which pattern rewrites are applied at the granularity of an individual operation. Extending, formalizing, and verifying the rewriting logic automatically would be an important next step \cite{DBLP:journals/scp/BravenboerKVV08,Meseguer:2010:TYR:1927806.1927809}.
On the backend side, \MLIR's DDR has an analogue to LLVM's instruction selection infrastructure, supporting extensible operations with multi-result patterns and specification as constraints \cite{Thier:2018:FFI:3178372.3179501}.

Numerous programming languages and models tackle hardware heterogeneity. Originally a homogeneous programming model, OpenMP added support for offloading tasks and parallel regions to accelerators \cite{OpenMP}, based on earlier proposals such as StarSs and OpenACC \cite{DBLP:journals/ijhpca/PlanasBAL09,OpenACC}. \Cpp AMP, HCC and SyCL leverage a conventional Clang/LLVM flow and modern \Cpp to provide a high-level abstraction for hardware acceleration~\cite{SyCL}. Unfortunately, all these examples very quickly lower high-level constructs to calls to a runtime execution environment, relying on pre-existing optimizations in the host language (typically \Cpp) to alleviate the abstraction penalty. Far fewer efforts target the heterogeneous compilation process itself. Parallel intermediate representations extending LLVM IR address part of the issue but traditionally focus on the homogeneous setting \cite{Khaldi:2015:LPI:2833157.2833158,Schardl:2017:TEF:3155284.3018758}. The most ambitious effort to date may be Liquid Metal \cite{Auerbach:2012:CRH:2228360.2228411}, with a co-designed Domain Specific Language (DSL) and compilation flow converting managed object semantics into static, vector or reconfigurable hardware; yet most of the effort in its Lime compiler reside in fitting round objects into square hardware (paraphrasing Kou and Palsberg \cite{Kou:2010:OFF:1869459.1869470}). \MLIR provides a direct embedding for high level languages embracing heterogeneity through extensible set of operations and types, while providing a common infrastructure for gradually lowering these constructs with maximal reuse of common components across the different targets.

Tackling language heterogeneity has been a long-term promise of metaprogramming systems, and of multistage programming in particular. Lightweight Modular Staging (LMS)~\cite{DBLP:journals/cacm/RompfO12} is a state of the art framework and runtime code generator, providing a library of core components for generating efficient code and embedding DSLs in Scala. Delite~\cite{DBLP:journals/tecs/SujeethBLRCOO14} promises dramatic productivity improvements for DSL developers, while supporting parallel and heterogeneous execution. We believe this approach is complementary to \MLIR, providing a higher-level of abstraction to embed DSLs and implement optimizations through generic metaprogramming constructs.

One step further up into the language syntax,
ANTLR~\cite{antlr} is among a class of parser generators that aim to make it
easy to develop a new compiler frontend. \MLIR currently does not have a general
parser generation, no AST construction or modeling functionality.
Combining \MLIR
with a system such as ANTLR could result in reusable compiler libraries
from user input through to code generation.

More narrowly construed by their application to machine learning, XLA~\cite{xla},
Glow~\cite{rotem2018glow} and TVM~\cite{auto-tvm}, address
similar heterogeneous compilation objectives. Yet these are rather specific code generation instances starting
from a graph abstraction and targeting multi-dimensional vector abstractions for
accelerators. All of these could leverage \MLIR as infrastructure, taking advantage of
the common functionality while using their current code generation strategies.
Similarly, the loop nest metaprogramming techniques from Halide~\cite{halide}
and
TVM~\cite{auto-tvm}, earlier loop nest
metaprogramming~\cite{uruk,10.1007/978-3-642-19595-2_10,DBLP:conf/cgo/BagneresZHB16,Cohen:2006:SPG:1161486.1161489},
and fully automatic flows such as PolyMage~\cite{mullapudi2015asplos}, Tensor
Comprehensions~\cite{tc}, Stripe~\cite{plaidml}, Diesel~\cite{Elango:2018:DDL:3211346.3211354}, Tiramisu \cite{Bag19} and their underlying polyhedral compilation
techniques~\cite{Fea92b,isl,bondhugula08pldi,ppcg} could
co-exist as different code
generation paths within an \MLIR-based compilation framework.
Serialization and interoperability formats, such as ONNX~\cite{onnx}, have a
different approach towards addressing the diversity of ML frontends by providing
a common set of ops that different frameworks could map on to. ONNX would be a
candidate as a dialect in \MLIR to which other ops could be lowered to and from.

\maybebreak
\section{Conclusion and Future Work}

We presented \MLIR, a flexible and extensible infrastructure for compiler
construction. This paper described \MLIR's concrete
design, demonstrated its applicability to a range of
important domains, and described a number of original research and engineering
implications.

Looking ahead, we are eager to see how established compiler communities
(e.g.\ the Clang C and \Cpp compiler) as well domain experts can benefit from the
introduction of higher level, language-specific IRs. We are also interested to
see if \MLIR enables new approaches to teaching the art of compiler and IR
design, and hope to see entirely new areas of research catalyzed or accelerated
by this infrastructure.

\paragraph{Future directions}

There are multiple future directions being pursued for \MLIR.
In the ML and HPC space, these include inferring efficient Op implementations from reference rank-polymorphic specifications with symbolic shapes. It also involves enabling a wider range of data structures (sparse, graphs) and program transformations, bringing together symbolic reasoning, such as automatic differentiation and algorithmic simplification, with more conventional data flow and control flow-based optimizations. Beyond ML and HPC, one may consider \MLIR's applicability to other related domains, such as secure compilation, safety-critical systems, data analytics and graph processing, relational query optimization, etc.

Returning to the world of general-purpose languages, an obvious missing front-end
is a C++ mid-level
representation derived from Clang. Say, a ``CIL'' similar in spirit to Swift's SIL and Rust's MIR, that would facilitate the optimization of
common C++ idioms that currently need to be reconstructed from lowered code (e.g.,
treating \lstinline{std::vector} as an array rather than pointer manipulation). Supporting garbage-collected languages, higher-order and polymorphic type systems with type inference in \MLIR are open challenges as well.

Exploring parallelism and concurrency constructs in LLVM has been difficult,
primarily as the changes required are invasive and not easily layered (e.g.,
injecting metadata and inspecting all passes to ensure metadata is propagated
while losing optimization opportunities as the level of abstraction is too low).
With \MLIR, parallel constructs
can be first-class operations, using regions and parallel idiom-specific verification. This would support
higher-level transformations before lowering to, e.g., LLVM where regular
transformations can be performed on already lowered code.

Beyond debugging and testing, textual form of IR is
also useful for education. Additional tooling to show the interaction of
optimizations in high-performance compilation could demystify compilers for new
students. IR design is an integral part of developing a new compiler or
optimization framework, but many undergraduate compiler curricula do not
cover IR design. \MLIR provides opportunities for new approaches to such 
lessons that could be explored.




\end{document}